\documentclass[twocolumn,preprintnumbers,amsmath,amssymb,prb]{revtex4}
\usepackage{graphicx} 
\graphicspath{{figs/}}
\usepackage{subfigure}
\usepackage{dcolumn}
\usepackage[usenames,dvipsnames]{color}
\usepackage{bm}

\def \be {\begin{equation}}
\def \ee {\end{equation}}
\def \ben {\begin{eqnarray}}
\def \een {\end{eqnarray}}

\begin{document}
\author{Daniel Montemayor}
\author{Eva Rivera}
\author{Seogjoo J. Jang}
\email{sjang@qc.cuny.edu}
\affiliation{Department of Chemistry and Biochemistry, Queens College, City University of New York, 65-30 Kissena Boulevard, Queens, New York 11367\footnote{mailing address}  \& PhD programs in Chemistry and Physics, and Initiative for the Theoretical Sciences, Graduate Center, City University of New York, 365 Fifth Avenue, New York, NY 10016}

\title{Computational Modeling of Exciton-bath Hamiltonians for LH2 and LH3 Complexes of Purple Photosynthetic Bacteria at Room Temperature} 
\date{Published in the {\it Journal of Physical Chemistry B} {\bf 122}, 3815-3825 (2018)}

\begin{abstract}
Light harvesting 2 (LH2) complex is a primary component of the photosynthetic unit of purple bacteria that is responsible for harvesting and relaying excitons.  The electronic absorption line shape of LH2 contains  two major bands at 800 nm  and 850 nm wavelength regions.   Under low light condition, some species of purple bacteria replace LH2 with LH3, a variant form with almost the same structure as the former but with distinctively different spectral features.  The major difference between the absorption line shapes of LH2 and LH3 is the shift of the 850 nm band of the former to a new 820 nm region.     The microscopic origin of this difference has been subject to some theoretical/computational investigations.  However, the genuine molecular level source of such difference is not clearly understood yet.  This work reports a comprehensive computational study of LH2 and LH3 complexes so as to  clarify different molecular level features of LH2 and LH3 complexes and to construct simple exciton-bath models with a common form.  All-atomistic molecular dynamics (MD) simulations of both LH2 and LH3 complexes provide detailed molecular level structural differences of BChls in the two complexes, in particular, in their patterns of hydrogen bonding (HB) and torsional angles of the acetyl group.  Time-dependent density functional theory calculation of the excitation energies of BChls for structures sampled from the MD simulations, suggests that the observed differences in HB and torsional angles  cannot fully account for the experimentally observed spectral shift of LH3.   Potential sources that can explain the actual spectral shift of LH3 are discussed, and their magnitudes are assessed through fitting of experimental line shapes.   These results demonstrate the feasibility of developing simple exciton-bath models for both LH2 and LH3, which  can be employed for large scale exciton quantum dynamics in their aggregates. 
\end{abstract}

\maketitle 
\noindent

\section{Introduction}
Purple bacteria have evolved highly specialized light-harvesting antenna units\cite{hu-qrb35,cogdell-qrb39} to efficiently transfer excitation energy to their reaction centers (RCs).  Under normal condition, they consist of two types of antenna complexes known as light harvesting 1 (LH1) and light harvesting 2 (LH2), both of which are cylindrically arranged trans-membrane pigment-protein complexes with bacteriochlorophyll (BChl) molecules as major pigments. While  LH1 is the larger of the two complexes and processes the penultimate exciton transfer step to the RC contained within itself, the primary functions of absorbing photons and delivering excitons over long distances are performed by the LH2 complexes.   Interestingly,  under low-light conditions, certain strains of purple bacteria can produce variant forms of LH2 called light harvesting 3 (LH3), which have different protein sequences but have similar structures as LH2.   

 Proteins constituting an LH3 complex can have heterogeneous compositions as in {\it {Rhodopseudomonas} ({\it Rps.}) {\it palustris}},\cite{brotosudarmo-bj97} consisting of both the original polypeptides of LH2 and alternate sequences.  They can also have homogeneous compositions as in {\it Rhodoblastus} ({\it R.}) {\it acidophilus} (formerly known as {\it Rhodopseudomonas} ({\it Rps.}) {\it acidophila}), consisting entirely of alternate sequences.  As the system becomes more stressed in its light condition, the overall proportion of the LH3 complexes increases.     This indicates that the variation from LH2 to LH3 is a natural adaptation with an implication for light harvesting efficiency.  Thus,  detailed molecular level understanding of the spectroscopic tuning mechanism has significant implication in elucidating nature's strategy to achieve optimal light harvesting functionality.
 
Both LH2 and LH3 have two distinct aggregates of BChl molecules that are respectively named after the wavelengths of exciton bands originating from the {\it $Q_y$} transitions of BChls.  The higher energy band at 800 nm wavelength called B800 is common for both LH2 and LH3, and is formed by one third of the BChls that are located at the periphery of the protein near the water-membrane interface.   On the other hand, the 850 nm region band in LH2 called B850, which are formed by two thirds of BChls forming two concentric rings (labeled as $\alpha$ and $\beta$) within the hydrophobic core of the protein, shifts to a region around 820 nm for the case of LH3, which is thus called B820.  This blue shift is expected to enhance the transfer rate of  excitons from the B800 band at the expense of narrowing the absorption range.\cite{ma-jpcb102}  It can also make the back transfer of excitons from LH1 less favorable,\cite{denium-bba1060} thus increasing the overall quantum yield.

\begin{figure}
\begin{center}
\includegraphics[width=\linewidth]{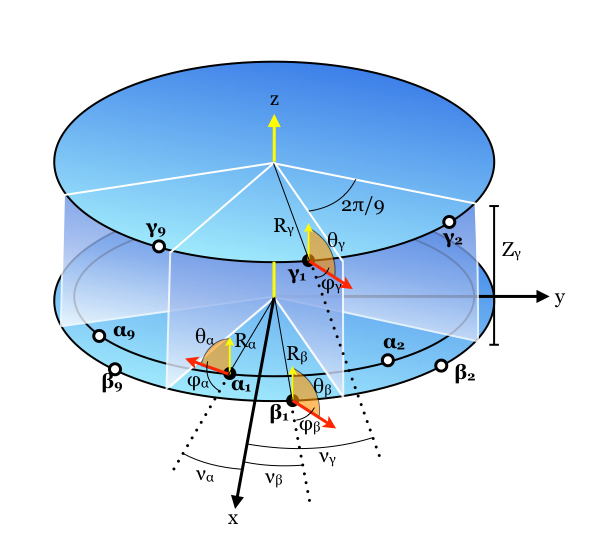}
\caption{Schematic of the 9-fold LH2 (LH3) complex.
The first protomer ($n=1$) and its 2 adjacent ones ($n=2, 9$), are shown appearing as wedges sliced into the cylindrical complex. 
BChl molecules appear as filled (for $n=1$) or open (for $n=2,9$) circles with bold labels $\alpha_n$, $\beta_n$, and $\gamma_n$ representing each site. 
Yellow arrows indicate axial $z$-directions.
The pigment radial distance from the $z$-axis is depicted by $R$. Red arrows represent BChl transition dipoles. Polar angle $\theta$, shown in orange, is the angle the transition dipole makes with the $z$-axis. The azimuthal angle $\nu$ is the angle the BChl position vectors of the first protomer make with the $x$-axis, while the azimuthal angle $\phi$ is the angle of the $xy$ projection of the transition dipole relative to the radial vector pointing to each BChl on the $xy$-plane. $Z_\gamma$ is the distance separating the plane of B800 BChls  from that of B850 (B820) BChls, which are defined to lie on $z=0$ plane. }
\label{fig:LH2schematic}
\end{center}
\end{figure}

For the case of {\it R. acidophilus}, which is the subject of this work, the X-ray crystallography data\cite{mcdermott-nature374,papiz-jmb326,mcluskey-biochemistry40} of LH2 and LH3  clearly exhibit similar tertiary structures, both consisting of 9-fold circular arrangements of $\alpha$-BChl and $\beta$-BChl holding polypeptides (see Fig. \ref{fig:LH2schematic}).    However,  the sequences of these  $\alpha$ and $\beta$ polypeptides are different for LH2 and LH3, with sequence identities of 68\%  and 74\% for $\alpha$ and $\beta$ respectively.  As a result, while both complexes have similar arrangements of BChls overall, subtle differences exist in the local environments of BChls. These differences appear to be responsible for the major 850 to 820 nm absorption band shift.  Clarifying their effects have been the subject of a number of previous studies,\cite{chmeliov-jpcb117,freiberg-cpl500,deruijter-cp341,zigmantas-pnas103,ritz-jpcb105} but no prevailing explanation is available yet.

 The native BChl binding pockets in LH2 or LH3 can accommodate large pigment nuclear rearrangements of peripheral functional groups of the tetrapyrrole-ring without significantly altering the pigment orientation. In fact, the entire pigment can be exchanged for another with radically altered functional groups, while conserving the orientation of the original pigment molecule. \cite{herek-bj78} An important factor contributing to the disparity in local environments between LH2 and LH3 results from variation in key protein residues participating in the hydrogen bonding (HB) with BChls.  Detailed investigation of the impact of such replacement is important in the context of general understanding of the role of HB, which have garnered attention in different contexts for several years now.\cite{cogdell-qrb39,jang-jpcl6,uyeda-biochemistry49,freiberg-bj103,higashi-jpcb118} 
 
In LH3, two protein residues responsible for HB in LH2 are both altered.  This  is similar in execution to an experiment conducted by Fowler and co-workers,\cite{fowler-nature355} where two B850 hydrogen bonding partners, $\alpha$-polypeptide residues Trp45 and Tyr44, were systematically mutated out of the sequence, resulting in a blue shift of the 850 nm band.  A single mutation of the Trp45 residue results in the shift of the 850 nm band to approximately 840 nm, while a double mutation of both Trp45 and Tyr44 shifts it to approximately 830 nm.  This qualitative resemblance to the LH3 line shape suggests the lack of HB as major mechanism for increasing the excitation energies of BChls. Indeed, our recent computational study\cite{jang-jpcl6} has revealed that the HB of BChl to either Tyr or Trp residue produces solvatochromic red shift of the BChl excitation energy by $\sim 500 {\rm \ cm^{-1}}$, supporting experimental results by Fowler {\it et al.}\cite{fowler-nature355}
 
In LH3, it is clear that the $\beta$-BChl of the B820 band does not have any proximate protein residue as HB partner. However,
despite the alternation of two protein residues responsible for HBs in LH3 as noted above, the crystal structure\cite{mcluskey-biochemistry40} suggests that the $\alpha$-BChl is still able to form an alternate HB with a nearby Tyr41.  Therefore, if HB alone is responsible for the excitation energy shift, then one would expect the LH3 line shape to resemble Fowler's single mutation spectrum.\cite{fowler-nature355} Instead, the line shape is more like that of the double mutated one.   This strongly suggests that there has to be another source for the blue shift of $\alpha$-BChl in LH3. One possibility is that the acetyl group rotation of $\alpha$-BChl in LH3, which is observed in the X-ray crystal structure and is promoted by the HB to Tyr41, may contribute to the blue shift.  

 In the crystal structure of LH3,\cite{mcluskey-biochemistry40} it has been observed that the acetyl group of the $\alpha$-BChl in the B820 unit is rotated by an additional $30^\circ - 40^\circ$ out of plane relative to that observed in the B850 unit.  This rotation can effectively remove a double bond from the BChl ring conjugation, resulting in blue shift of the excitation energy.   Gas phase ZINDO calculations\cite{gudowska-nowak-jpc94} demonstrated a clear dependence of the BChl excitation energy on the rotation of the acetyl group, resulting in maximal blue shifts in excess of 300 ${\rm cm^{-1}}$.  However, the quantitative reliability of this calculation is questionable considering its semi-empirical nature based on old set of experimental data in solution phase, let alone the possibility that the actual distortion of the acetyl group in membrane environment may be different from that of the crystal structure data.  In fact, a similar variation of the acetyl group rotation was also found among the BChl pigments of the FMO complex with density functional theory (DFT) calculations, \cite{muh-pnas104} but the magnitude of the energy difference was only about 100 ${\rm cm^{-1}}$ for comparable values of the torsional angle. 

While the rotation of the acetyl group as described above may serve as a possible tuning mechanism of excitation energy, its actual extent {\it in vivo} and the resulting change in the excitation energy of $\alpha$-BChls in LH3 has not been well established. A reasonable way to examine this issue is (i) to sample structures of BChls, through molecular dynamics (MD) simulation of the membrane-bound complex under biological condition, and (ii) to conduct comparative  {\it ab initio} calculation of  the excitation energies of B820-BChls and B850-BChls for the sampled structures.

In this work we present a comprehensive study of LH2 and LH3 complexes, integrating all-atomistic MD simulation of the full membrane-bound system under biological condition and time dependent (TD) DFT calculation of $Q_y$ excitation energies of BChls that have been optimized together with proximate protein residues by DFT calculations.  Compilation and analysis of these data provide new insights into the potential molecular level factors that differentiate spectroscopic properties of LH2 and LH3, and also lead to compact forms of exciton-bath Hamiltonians that can be easily employed for the modeling of spectroscopic data and larger scale exciton dynamics simulation.

This paper is organized as follows.  Section II provides description of all-atomistic MD simulations for both LH2 and LH3 in membrane environments.  Section III provides a coarse-grained exciton-bath model that can be used commonly for both LH2 and LH3, and also offers the best set of parameters based on simulations.  Section IV provides detailed analysis of the excitation energies of BChls.  Section V discusses the implication of our computational data, and offers models that can best explain the ensemble absorptions spectroscopies of both LH2 and LH3 complexes. Section VI summarizes main outcomes and implications of our study.

\section{All-atomistic molecular dynamics simulation}
\label{MD}
MD simulations of the LH2 and LH3 complexes from {\it R. acidophilus}, with pdb codes 1NKZ \cite{papiz-jmb326} and 1IJD \cite{mcluskey-biochemistry40} respectively, were performed under biological condition. These include the 9-fold pigment-protein complexes  in membrane and explicit water solvent under the condition of standard temperature and pressure. Preparation of the computational models was made through the following multiple steps:  1) parameterization of the molecular mechanics (MM) force field; 2) application of crystallographic transformations on the pdb coordinates with appropriate adding/pruning of atoms and subsequent minimization; 3) insertion of the minimized complex into a pre-equilibrated membrane structure followed by charge neutralization; 4) a final minimization and equilibration of the model in preparation for the production run.

The MM force field was parameterized using well-documented sources that are available.  	   
The CHARMM27 parameter set\cite{mackerell-jpcb102} was used for all standard residues.  Parameters for the BChl molecules were taken from Schulten and coworkers.\cite{mackerell-jpcb102} Partial charges for the BChl molecules used are reported elsewhere.\cite{rivera-jpcb117}  The parameters and partial charges of the carotenoid molecules, rhodopin glucoside (RG1 in LH2) and rhodopinal glucoside (RPA in LH3), were determined using the appropriate atom types given in the CGenFF force field.\cite{vanommeslaeghe-jcc31} This parameterization approach is common particularly for modeling light harvesting complexes that have BChls. \cite{olbrich-jpcb114,rivera-jpcb117,jang-jpcl6,vandervegte-jpcb119}

\begin{figure}
(a)\makebox[1.6in]{ }(b)\makebox[1.4in]{ } \\
\includegraphics[width=1.3in]{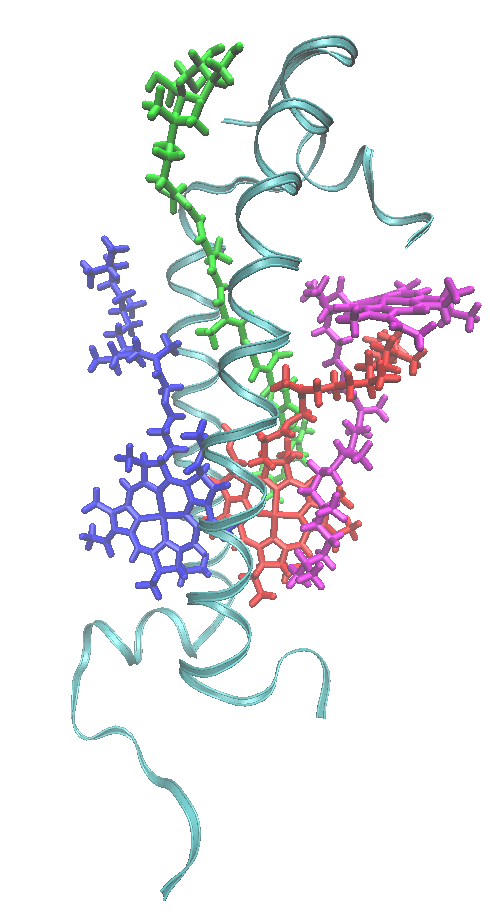} \includegraphics[width=1.7in]{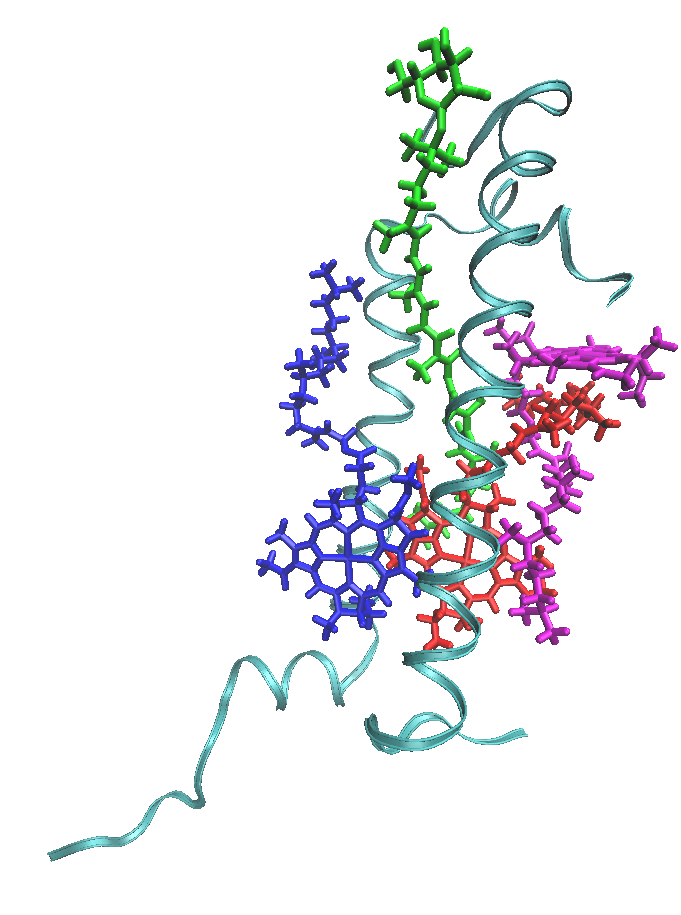}
\caption{Snapshots of the symmetry units for equilibrated (a) LH2 and (b) LH3 complexes.}
\label{fig:snap-shot}
\end{figure}

All atom coordinates of the minimized complex were derived from the pdb structure. 
The second set of RG1 molecules in the LH2 complex were removed from the structure because their presence is under debate\cite{gall-febsl580} and also in order to keep the concentrations of carotenoids consistent between LH2 and LH3.  Next, appropriate crystallographic transformations were applied to the pdb coordinates to generate the complete 9-fold LH2 and LH3 structures.  Using VMD\cite{humphrey-jmg14.1} in conjunction with NAMD,\cite{philips-jcc26} missing atoms within the structure were added, and the complex was then minimized for 500 steps with a time step of $1\ {\rm fs}$.

Following minimization, the complexes were then inserted into a pre-equilibrated lipid bilayer\cite{zhao-bj92,zhao-biochimie90} consisting of 75\% POPE lipids, 25\% POPG lipids, a small layer of TIP3P\cite{jorgensen-jcp79} waters, and sodium counter-ions.  Overlapping lipid molecules within 0.8 {\AA} of any complex atom were deleted.  Next, the number of sodium ions was adjusted to produce a charge neutral system.  The system was then solvated with an additional layer of TIP3P waters along the axial direction of the complex to prevent interaction between periodic images.  The final models contained approximately 150,000 atoms  with simulation box dimensions approximately 115 {\AA} $\times$ 114 {\AA} $\times$ 127 {\AA}.
  
\begin{figure}
\includegraphics[width=\linewidth]{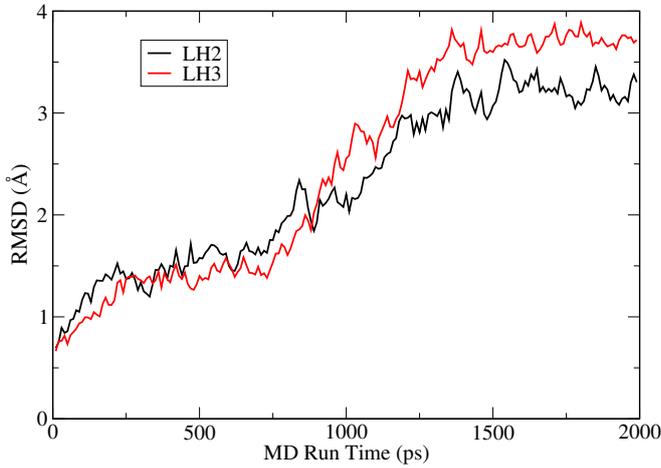}
\caption{Time dependent root mean square displacements (RMSDs) of protein backbones of LH2 and LH3 complexes during the second phase of MD simulations. }
\label{fig:rmsd}
\end{figure}
\noindent

In preparation for the production run, the LH2 and LH3 systems were then minimized for 150 steps  and equilibrated for 10 ns at 300 K in the NPT ensemble using a 1 fs time step.  Periodic boundary conditions were employed and particle mesh Ewald (PME)\cite{darden-structure7} method was used to account for the long-range electrostatics.   Figure \ref{fig:snap-shot} provides snapshots of the symmetry units of LH2 and LH3 complexes after the equilibration.  Following the equilibration period, two simulation phases were conducted for data production. The first phase consisted of a 60 ps production run with the system coordinates recorded every 5 fs.  This first phase production was used to compute the bath spectral density for each BChl molecule, which will be described in more detail in the next section.  The second phase of the production run was a continuation of the simulation for an additional 2 ns with coordinates recorded every 1 ps to gather further configurational statistics.  Figure \ref{fig:rmsd} shows running values of root mean square displacements (RMSDs) during this phase.  The saturation of RMSDs indicates that reasonably sufficient sampling of configurations has been made during the 2 ns time interval. This was used to calculate geometric parameters entering the coarse-grained exciton model.   It is important to note that the structures have already been equilibrated before the 2 ns samplings shown in Fig. 3.  The slow approach to steady state values in the RMSDs during these simulations indicate that more than 1 ns is required to sample representative structures even after equilibration.

\section{Exciton-bath model}
\subsection{Structures}
The aggregates of BChls realized in both LH2 and LH3 can be represented by a common geometric model extending those developed previously.\cite{jang-jpcb105,jang-jcp118-2,jang-jpcb111} 
Both can be viewed as cylindrical complexes belonging to the ${\rm C_9}$ point group, where each symmetry unit contains one BChl pertaining to the B800 band and a pair of BChls ($\alpha$ and $\beta$) constituting the B850 (B820) band for LH2 (LH3).   To simplify the notation, from now on, we label the BChl constituting the B800 band as $\gamma$.

$\gamma$-BChls are located about $16\ {\rm \AA}$ in the axial $z$ direction relative to the $\alpha$- and $\beta$-BChls that lie on the $xy$ plane.
The $x$-axis is defined as bisecting the $\alpha$- and $\beta$-BChls of the first symmetry unit. These symmetry units are also often referred to as protomers, a term we shall use interchangeably throughout the text. Due to the ${\rm C_9}$ symmetry, the position and orientation of BChls in other symmetry units can be generated through rotation around the $z$-axis by $2\pi n/9$, with $n=1,\cdots, 8$.  
A schematic of this model can be seen in Fig. \ref{fig:LH2schematic}, where cylindrical coordinates ($R$, $\nu$, $Z$) represent the position of a BChl, represented by its Mg center.  The spherical coordinates ($\mu$, $\theta$, $\phi$) in this figure represent the $Q_y$ transition dipole moment vector of BChl.

In cartesian coordinates, the pigment position and transition dipole moment are respectively
\begin{equation}
\label{Rxyz}
{\bf R}_{s,n}=\left (
\begin{array}{c}
R_{s}\cos\left [2\pi (n-1)/9+\nu_{s}\right ] \\
R_{s}\sin\left  [2\pi (n-1)/9+\nu_{s}\right ] \\
Z_s  \end{array}\right )
\end{equation}
and
\begin{equation}
\label{muxyz}
\boldsymbol{\mu}_{s,n}=\mu_s
\left(
\begin{array}{c}
\sin\theta_{s}\cos[2\pi (n-1)/9+\nu_{s}+\phi_{s}]\\
\sin\theta_{s}\sin[2\pi (n-1)/9+\nu_{s}+\phi_{s}]\\
\cos\theta_{s}
\end{array} \right)
\end{equation}
where $s=\alpha,\beta,\gamma$ and $n=1,\cdots, 9$.

The second phase of the MD simulation after equilibration has been used to determine the geometric parameters in Eqs. (\ref{Rxyz}) and (\ref{muxyz}), by taking the average of coordinates of BChls and their transition dipoles from the simulation.   For the transition dipole, we have used  the point transition charge approximation (TrEsp method) developed and parametrized by Renger and co-workers,\cite{madjet-jpcb110,adolphs-pr95,renger-pr102}  which assigns atomic transition charges $q^T$ centered on the pigment nuclear coordinates to approximate the molecular transition dipole $\langle \psi_e|\hat\mu|\psi_g\rangle$.   The resulting values of geometric parameters and transition dipoles are summarized in Table \ref{geometry}.   Comparison of these with those extracted from the X-ray crystal structure data\cite{krueger-jpcb102,jang-jcp118-2} show that the differences are relatively small and subtle.

MD simulation results demonstrate that LH2 and LH3 complexes have nearly identical tertiary structures and pigment arrangements.  Analyses of the second phase MD simulations reveal this geometric similarity in terms of the model parameters illustrated in Fig. \ref{fig:LH2schematic} and summarized in Table \ref{geometry}. For both complexes, these pigment coordinates are tightly restrained during the 2 ns simulation runs. The most significant geometric difference between the two complexes is in the molecular orientation of their respective B850 and B820 BChls. The azimuthal coordinate $\phi_\beta$ is displaced between the two complexes by about $5^\circ$ while, to a lesser degree, the zenith angles $\theta_\alpha$ and $\theta_\beta$ of the two complexes differ by $2-3^\circ$.  While these changes do not have significant effects on most values of electronic coupling constants, they result in substantial reduction of the nearest-neighbor electronic couplings between $\alpha$- and $\beta$-BChls from  adjacent protomers, $J_{\alpha\beta}(1)$, by about half (see the second row of Table \ref{MDparams}).  
\begin{table}
   \centering
   \caption{Geometric parameters appearing in Fig. \ref{fig:LH2schematic}.  Values are derived from the second phase MD simulations and represent averages over the sampled time dependent coordinates and over the 9 pigments of the same type.  Transition dipole moments are in Debye.} 
   \begin{tabular}{@{} cccc @{}} 
     \hline\hline  
 & LH2 & LH3 & rel. diff. \% \\
 \hline
     $R_\alpha$ & 26.0$\pm$0.1 {\AA} & 25.6$\pm$0.1 {\AA}&1.54\\
     $R_\beta$ & 27.5$\pm$0.1 {\AA} &27.4$\pm$0.1 {\AA}&0.36\\
     $R_\gamma$ & 32.1$\pm$0.1 {\AA} & 31.2$\pm$0.1 {\AA}&2.80\\
     $Z_\gamma$ & 16.5$\pm$0.1 {\AA} & 16.4$\pm$0.1 {\AA}&0.61\\
     \hline
     $\theta_\alpha$ & 96.5$\pm$0.7$^\circ$ & 93.7$\pm$0.6$^\circ$&2.90\\
     $\theta_\beta$ & 97.3$\pm$0.9$^\circ$ & 98.9$\pm$0.9$^\circ$&1.64\\
     $\theta_\gamma$ & 98.2$\pm$0.8$^\circ$ & 99.0$\pm$0.8$^\circ$&0.81\\
     \hline
     $\nu_\alpha$ &-10.20$\pm$0.07$^\circ$ &-10.24$\pm$0.06$^\circ$&0.39\\
     $\nu_\beta$ &10.20$\pm$0.07$^\circ$ &10.24$\pm$0.06$^\circ$&0.39\\
     $\nu_\gamma$ & 23.3$\pm$0.3$^\circ$ &23.8$\pm$0.2$^\circ$&0.21\\
     \hline
     $\phi_\alpha$ & -106.6$\pm$0.6$^\circ$ &  -107.1$\pm$0.6$^\circ$&0.47\\
     $\phi_\beta$ &  60.0$\pm$0.9$^\circ$ & 55.4$\pm$1.0$^\circ$&7.67\\
     $\phi_\gamma$ &  63.7$\pm$0.8$^\circ$ & 60.8$\pm$0.7$^\circ$&4.55\\
     \hline
     $\mu_\alpha$ & 6.41$\pm$0.02 ${\rm D}$& 6.34$\pm$0.02 ${\rm D}$&1.09\\
     $\mu_\beta$ & 6.30$\pm$0.02 ${\rm D}$&  6.36$\pm$0.02 ${\rm D}$&0.95\\
     $\mu_\gamma$ & 6.48$\pm$0.01 ${\rm D}$& 6.46$\pm$0.02 ${\rm D}$&0.31\\
      \hline
      \hline
   \end{tabular}
   \label{geometry}
\end{table}

\subsection{Exciton-bath Hamiltonian}
The B800 and B850 (B820) bands of LH2 (LH3) are single exciton states formed by $Q_y$ transitions of BChls, and are known (expected) to be described well by the Hamiltonian of Frenkel exciton\cite{frenkel-pr37,knox,davydov,chernyak-jcp105} and linearly coupled bath of harmonic oscillators.  Extending the notations adopted before,\cite{jang-jcp118-2,jang-jpcb111,jang-jpcb115,kumar-jcp138,jang-jpcl6} we here introduce the following total exciton-bath Hamiltonian:
\begin{eqnarray}
H_T=\epsilon_g|g\rangle\langle g|+H_e^0+\delta H_e+H_b+H_{eb}\ , \label{eq:h_t}
\end{eqnarray}  
where $|g\rangle$ is the ground electronic state with energy $\epsilon_g$, $H_e^0$ is the average exciton Hamiltonian for LH2 (LH3) with perfect ${\rm C_9}$ symmetry, and $\delta H_e$ represents the disorder in the exciton Hamiltonian.  $H_e^0$ can be expressed  as follows:
\begin{eqnarray}
 H_e^0&=&\sum_{n=1}^9 \left \{ \epsilon_\alpha |\alpha_n\rangle\langle \alpha_n|+\epsilon_\beta |\beta_n\rangle \langle \beta_n|+\epsilon_\gamma|\gamma_n\rangle \langle \gamma_n| \right . \nonumber \\
 &&\hspace{.2in}+J_{\alpha\beta}(0)(|\alpha_n\rangle\langle \beta_n|+|\beta_n\rangle\langle \alpha_n|)  \nonumber \\
 &&\hspace{.2in}+J_{\alpha\gamma}(0)(|\alpha_n\rangle\langle \gamma_n|+|\gamma_n\rangle\langle \alpha_n|) \nonumber \\
 &&\hspace{.2in}\left . +J_{\beta\gamma}(0)(|\beta_n\rangle\langle \gamma_n|+|\gamma_n\rangle\langle \beta_n|) \right\}\nonumber \\
 &&+\sum_{n=1}^9\sum_{m\neq n}^9 \sum_{s,s'=\alpha,\beta,\gamma} J_{ss'}(n-m)|s_n\rangle\langle s'_m|\ , \label{eq:h_e0}
 \end{eqnarray}
where $|s_n\rangle$, with  $s=\alpha$, $\beta$, or $\gamma$, represents the excited state where only $s_n$-BChl is excited with its energy $\epsilon_s$, and $J_{ss'}(n-m)$ is the electronic coupling between $|s_n\rangle$ and $|s'_m\rangle$.  
The term $\delta H_e$ in Eq. (\ref{eq:h_t}) represents the disorder Hamiltonian and can be assumed to be diagonal in the site excitation basis as follows: 
\begin{eqnarray}
\delta H_e=\delta \epsilon_g|g\rangle\langle g|+\sum_{n=1}^9 \sum_{s=\alpha,\beta,\gamma} \delta \epsilon_{s_n}|s_n\rangle\langle s_n|\ ,  \label{eq:delta_he}
\end{eqnarray}
where $\delta \epsilon_g$ is the disorder in the ground electronic state, which represents deviation of the sum of all the energies of  BChls in their ground electronic state from the average value, and $\delta \epsilon_{s_n}$ is the disorder in the excited state energy of a particular $s_n$-BChl.      Thus, the excitation energy of $s_n$-BChl corresponds to 
\be
E_{s_n}^{ex}=\epsilon_s+\delta \epsilon_{s_n}-\epsilon_g-\delta \epsilon_g \ .  
\ee
Without losing generality, we assume that $\epsilon_g=0$.  Thus, $\epsilon_s$ is assumed to be equal to the average excitation energy of  $s$-BChl.   
The assumption behind Eq. (\ref{eq:delta_he}), is that the disorder exists only in the excitation energies of BChls and not in their electronic couplings.  This simplification can be justified by the fact that the disorder in the off-diagonal components (electronic couplings) are relatively small (see Table \ref{MDparams}) compared to those in site excitation energies.   

The disorder in site energies may come from many different sources. Under the assumption that the central limit theorem is applicable, the disorder in site excitation energies can be modeled as having Gaussian distribution.   Experimental ensemble line shapes for LH2 suggest that the standard deviations of the Gaussian distribution are on the order of $200{\rm \ cm^{-1}}$ for the B850 unit.  Similar values are expected for the B820 unit  of LH3.  Fine tuning of actual values of the disorder can be made during the modeling of ensemble spectroscopic data, once the zeroth order  Hamiltonian $H_e^0$ becomes well established.

\begin{table}[htbp]
   \centering
   \caption{Electronic coupling constants and reorganization energies calculated from second phase MD simulations.  All values are in ${\rm cm^{-1}}$.}
\begin{tabular}{c|c|c|c}
     \hline\hline
     Parameter& LH2 & LH3& rel. diff. \%\\
     \hline 
         $J_{\alpha\beta}(0)$&$244.74\pm4.97$&$238.74\pm5.11$&2.45\\
     $J_{\alpha \beta}(1)$&$139.75\pm10.22$&$74.13\pm10.69$&46.96\\
      \hline 
     $J_{\alpha\alpha}(1)$&$-59.07\pm1.01$&$-61.21\pm1.01$&3.62\\
     $J_{\beta\beta}(1)$&$-29.27\pm1.61$&$-21.54\pm1.70$&26.41\\
     \hline 
     $J_{\beta\alpha}(1)$&$13.83\pm0.17$&$13.52\pm0.18$&2.24\\
     $J_{\alpha\beta}(2)$&$12.66\pm0.35$&$10.83\pm0.40$&14.45\\
     \hline
     $J_{\gamma\gamma}(1)$&$-24.46\pm0.77$&$-23.45\pm0.75$&4.13\\
     $J_{\alpha\gamma} (1)$&$28.29\pm0.59$&$26.58\pm0.52$&6.04\\
     $J_{\beta\gamma} (1)$&$2.63\pm0.61$&$3.31\pm0.54$&25.86\\
     $ J_{\gamma\alpha}(0)$&$-12.31\pm0.42$&$-12.17\pm0.37$&1.14\\
     $ J_{\gamma \beta}(0)$&$-2.68\pm0.83$&$0.88\pm0.77$&67.16\\
     \hline
     $\lambda_{\alpha}$ & 23.9 & 27.1&13.39\\ 
     $\lambda_{\beta}$ & 28.4 & 26.4&7.04\\ 
     $\lambda_{\gamma}$ & 73.3 & 72.4&1.23\\ 
      \hline
      \hline
   \end{tabular}
   \label{MDparams}
\end{table}

All the terms constituting $H_e^0$, Eq. (\ref{eq:h_e0}), can in principle be calculated by {\it ab initio} methods.\cite{cupellini-jpcb120,anda-jctc12,sgatta-jacs139,sisto-pccp19}  Full quantum consideration of the entire LH2 and LH3 complex may be ideal, but the efficiency and accuracy of such calculations do not seem to be satisfactory yet.  Thus, explicit quantum calculation of only BChls or with proximate protein residues, while treating the rest in terms of an appropriate solvation model,\cite{cupellini-jpcb120,sgatta-jacs139} seems like the only viable approach at present. 

In our recent study,\cite{jang-jpcl6} we have employed the TD-DFT method to calculate excitation energies of BChls along with their proximate protein residues within about ${\rm 5\ \AA}$ distance, without considering other solvation effect of the surrounding protein environments.  The purpose of these calculations was to estimate the effect of HB on the excitation energy of BChl, rather than reproducing exact values of excitation energies, which requires higher level calculation method and full consideration of solvation effects.   A similar approach is used in this work.  Detailed account of TD-DFT calculation results will be provided in the next section. 

For the calculation of electronic couplings, $J_{ss'}(n-m)$ in Eq. (\ref{eq:h_e0}),  various methods invoking different levels of approximation\cite{krueger-jpcb102,tretiak-jpcb104,madjet-jpcb110,neugebauer-cpc10} can be used. Since our objective here is to calculate the values averaged over the MD trajectories, an efficient method is needed.  For this, we have employed the TrEsp method developed by Renger and co-workers.\cite{madjet-jpcb110,adolphs-pr95,renger-pr102} The TrEsp method assigns atomic transition charges $q^T$'s centered on the pigment nuclear coordinates to approximate the molecular transition dipole moment, and calculates electronic coupling between excitations of pigment molecules by assuming only Coulomb interactions as follows: 
\begin{equation}
\label{Delta}
J_{ss'}(n-m)=\frac{f}{4\pi\epsilon_0}\sum_{l\in s_n, l^\prime\in s'_m}^L\frac{q^T_l q^T_{l^\prime}}{|{\bf R}_{l,l^\prime}|}.
\end{equation}
Here, $l$ and $l^\prime$ index over the $L$ nuclei in pigments $s_n$ and $s'_m$ respectively, $|{\bf R}_{l,l^\prime}|$ is the distance separating nuclei $l$ and $l^\prime$, which can be obtained directly from the MD simulation data.  Values of $q^T_l$'s are reported elsewhere,\cite{madjet-jpcb110} and $f$ is the effective dielectric screening factor.   We use the value of $f=.55$, which  is chosen to approximate previously reported electronic coupling between LH2 sites calculated by transition density cube method.\cite{krueger-jpcb102} This value of $f$ is in accord with previous calculations in similar environments,\cite{scholes-jpcb111,olbrich-jpcb114}
and reproduces the reported value\cite{alden-jpcb101,knox-pp7,pullerits-jpcb101,scherz-bba766} of the magnitude of BChl transition dipole magnitude calculated by
\begin{equation}
\label{MM}
\mu_{s,n}^2 =\frac{f}{4\pi\epsilon_0}\sum_{l, l'\in s_n, l}^L q^T_l q^T_{l^\prime} {\bf R}_{l}\cdot {\bf R}_{l'}.
\end{equation}
Table \ref{MDparams} provides all the major electronic coupling constants calculated as described above and averaged over the MD trajectories.   Error estimates based on standard deviations resulting from different configurations along the MD trajectories are also shown.  \vspace{.3in}\\

A few important points in regard to the data in Table \ref{MDparams} are worth mentioning here.  Comparison of the values for LH2 complex with other data available,\cite{krueger-jpcb102,jang-jcp118-2,cupellini-jpcb120,sgatta-jacs139} shows that the electronic couplings averaged over the trajectories of MD simulation are in general smaller than those calculated for fixed X-ray crystal structure.  This is consistent with the results of recent calculation based on higher level methods.\cite{cupellini-jpcb120}  Although our values are somewhat smaller than the values obtained by Cupellini {\it et al.},\cite{cupellini-jpcb120} this is not expected to cause significant difference in the ensemble line shape as long as appropriate level of disorder is chosen for each case.

Finally, the bath and exciton-bath Hamiltonians in Eq. (\ref{eq:h_t}) represent all the dynamic response to the electronic excitation of BChls, which include intramolecular vibrational modes and response of the surrounding protein environments.   We here employ the conventional approximation of modeling such response by linearly coupled set of harmonic oscillators as follows:
\ben
&&H_b=\sum_j\hbar\omega_j\left (b_j^\dagger b_j+\frac{1}{2}\right ) \ ,\\
&&H_{eb}=\sum_j\sum_{n=1}^9\sum_{s=\alpha}^\gamma \hbar \omega_jg_{sn,j}(b_j+b_j^\dagger)|s_n\rangle\langle s_n| \ .
\een
Within this model,  the interaction of each $s_n$-BChl excitation with the bath can be characterized by the following spectral density:\cite{leggett-rmp59,weiss} 
\be
{\mathcal J}_{s_n}(\omega)=\pi\hbar\sum_j g_{sn,j}^2\omega_j^2 \delta(\omega-\omega_j) \ . \label{eq:spd_sn}
\ee 
Alternatively, in the coordinate representation of the bath oscillators, $\hbar\omega_j g_{sn,j}(b_j+b_j^\dagger)=c_{sn,j}Q_j$, where $Q_j$ is the mass-weighted $j$th normal mode coordinate and $c_{sn,j}=\sqrt{2\hbar} \omega_j^{3/2} g_{sn,j}$.  Therefore, the spectral density of Eq. (\ref{eq:spd_sn}) can also be expressed in terms of $c_{sn,j}$ as follows:\cite{rivera-jpcb117}  ${\mathcal J}_{s_n}(\omega)=(\pi/2)\sum_j (c_{sn,j}^2/\omega_j) \delta(\omega-\omega_j)$.   For this spectral density, the reorganization energy of the bath for the excitation of $s_n$-BChl is defined as
\be
\lambda_{s_n}=\frac{1}{\pi} \int_0^\infty d\omega \frac{{\mathcal J}_{s_n}(\omega)}{\omega}\ .
\ee

Given that the linearly displaced harmonic oscillator bath model provides realistic representation of the environmental response upon excitation of each BChl, the energy gap correlation function for the $s_n$-BChl in the classical limit can be expressed as 
\ben
\label{acf}
C_{E,s_n}^{cl}(t)&=&\langle \Delta E_{s_n}(0)\Delta E_{s_n} (t)\rangle_{cl}\nonumber \\
&=&\sum_{j}c_{sn,j}^2 \langle Q_j(0)Q_j(t)\rangle_{cl} \nonumber \\
&=&\frac{2}{\beta\pi}\int_0^\infty d\omega {\mathcal J}_{s_n}(\omega)\frac{\cos (\omega t)}{\omega}\ ,
\een
where $\langle \cdots\rangle_{cl}$ represents an average over the equilibrium classical ensemble.
Taking cosine transform of the above equation leads to 
\be 
{\mathcal J}_{s_n}(\omega)=\beta \omega \int_0^\infty dt \ C_{E,s_n}^{cl}(t) \cos (\omega t)\ .
\ee
The classical energy gap correlation function $C_{E,s_n}^{cl}(t)$ can be determined directly from the classical MD simulation.  Following the approach used before,\cite{rivera-jpcb117}  we here calculate the correlation of energy gaps originating from Coulomb interactions modeled at the level of the TrEsp method.\cite{renger-pr102,renger-pccp15} 

Partial atomic charges for the ground and excited electronic states were chosen to approximate their respective electronic densities in a manner analogous to the approximation for the transition dipole. The energy gap accompanying the excitation of $s_n$-BChl is approximated by the difference in the electrostatic interactions of its ground and excited state partial atomic charges with the protein partial atomic charges used in the MM force field as follows:
\begin{equation}
\label{deltaE}
\Delta E_{s_n}(t)=\frac{f}{4\pi\epsilon_0}\sum_{l\in s_n}\sum_{k} \frac{(q^e_l-q^g_l)q_k}{|{\bf r}_l(t)-{\bf R}_k(t)|} \ , 
\end{equation}
where  $l$ indexes over the nuclei of $s_n$-BChl, $k$ indexes over the nuclei of the protein environments, $q^g_l$ and $q^e_l$ respectively are the ground and excited state partial atomic charges of BChl as reported by Renger and co-workers,\cite{madjet-jpcb110} $q_k$'s are the partial charges\cite{mackerell-jpcb102,rivera-jpcb117} of the protein environments used in the MD forcefield calculation, and $|{\bf r}_l(t)-{\bf R}_k(t)|$ is the distance between the pigment and the environmental charges.  

Let alone the approximation used in the calculation of Columbo interactions, the above method to evaluate the bath spectral density ignores the effects of intramolecular vibration and the van der Waals interaction between BChl and proximate protein residues.   Therefore, further augmentation of our spectral densities reported here is necessary in order to construct complete ones reflecting all the exciton-bath couplings. Nonetheless, comparison of these spectral densities for different types of BChls in LH2 and LH3 helps understand the effects of different local protein environments. 

\begin{figure}[htbp]
\begin{center}
\ \vspace{.2in}\\
\includegraphics[width=\linewidth]{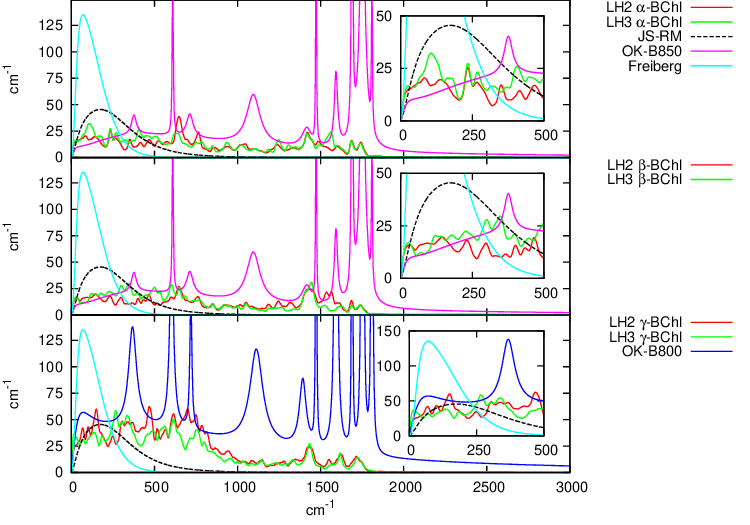}
\caption{Calculated spectral densities (divided by $\pi\hbar$) in the unit of ${\rm cm^{-1}}$ for  $Q_y$ excitation of BChls for LH2 (red) and LH3 (green) averaged over the 9 sites. The top panel is for $\alpha$, the middle panel for $\beta$, and the bottom is for $\gamma$-BChl.  Also compared are the spectral density by Jang {\it et al.}\cite{jang-jpcb111} based on that of Renger and Marcus,\cite{renger-jcp116} Olbrich and Kleinekath\"{o}fer,\cite{olbrich-jpcb114} and a recent one by Freiberg and coworkers.\cite{pajusalu-cpc12}  Insets show close-ups of regions below $500\ {\rm cm^{-1}}$. }
\label{specden}
\end{center}
\end{figure}

Figure \ref{specden} shows the spectral densities calculated from the first phase MD simulations and also compare with other known spectral densities for LH2 complex.\cite{jang-jpcb111,olbrich-jpcb114,pajusalu-cpc12} 
Overall, the spectral densities of corresponding pigment types are remarkably similar between LH2 and LH3.  The disparity in hydrophilicity between the environments of $\gamma$-BChl (B800 band)  and the $\alpha$,$\beta$-BChls (B850 or B820 band) is apparent.
Qualitatively, the spectral density for $\gamma$-BChl demonstrates pronounced density in the region below $1000\ {\rm cm^{-1}}$ when compared with those for $\alpha$- and $\beta$-BChls in both LH2 and LH3. This is due to the higher density of nearby charged and polar residues whose motion enhance site energy fluctuations. 

As reflected from the reorganization energy data in Table \ref{MDparams}, our simulation suggests that the excitation-bath coupling for $\gamma$-BChl appears to be significantly larger than those for $\alpha$- and $\beta$-BChls. However, these estimates are based on the assumption that the local dielectric constant around $\gamma$-BChl is the same as that for $\alpha$- and $\beta$-BChls. In reality, considering that the environment around the B800 band is more hydrophilic, it is expected that the effective dielectric constant entering the Coulomb interaction for $\gamma$-BChl is larger than those for $\alpha$- and $\beta$-BChls, resulting in reduction of actual Coulomb interactions.   In an earlier work,\cite{jang-jpcb111} an effective screening factor was used to account for this effect, which resulted in reasonable fitting of spectral line shape for the B800 band.    

 The most significant difference in the spectral densities for $\alpha$- and $\beta$-BChls of LH2 and LH3 complexes can be seen in the region below $500\ {\rm cm^{-1}}$.  For LH3, couplings to the bath are somewhat more pronounced in this region than those for LH2. The only region where there is stronger interaction for LH2 is for $\alpha$-BChl around $650\ {\rm cm^{-1}}$, the  source of which is not fully understood yet.  Despite some different characteristics as noted above, we find that the spectral densities of LH2 and LH3 obtained from MD simulation are similar overall.  Thus, they are not expected to be the source of different spectral features of LH2 and LH3 complexes.

\section{Excitation energies of BChls}

\subsection{Effect of hydrogen bonding}
The X-ray crystal structure\cite{papiz-jmb326} of LH2 shows that both $\alpha$- and $\beta$-BChls form HBs with protein residues.  Our MD simulations have  also demonstrated\cite{jang-jpcl6} that this form, with both native hydrogen bonds intact, is indeed most prevalent.  On the other hand, the crystal structure\cite{mcluskey-biochemistry40} of LH3 complex indicates that the $\beta$-BChl has no HB partner while the $\alpha$-BChl does, albeit to a different residue in the $\alpha$-polypeptide.  Similarly, the MD simulation of the LH3 complex reported here confirms that the state in which only the $\alpha$-BChl forms HB can be found dominantly. For a detailed description of how HB states and their prevalence is calculated, we refer the reader to our previous work.\cite{jang-jpcl6} 

Considering the high degree of probability for finding the double and single HB cases in LH2 and LH3 respectively, we can assume that practically these are the only two cases that exist for those complexes.  
Therefore, based on our earlier work,\cite{jang-jpcl6} we expect that the $\beta$-BChl of LH3 has higher excitation energy than that of LH2 by about ${\rm 500\ cm^{-1}}$.   In addition, the lack of HB also
gives rise to an apparent change in the azimuthal angle of the $\beta$-BChl, which in turn affects its nearest neighbor electronic coupling to $\alpha$-BChl, as can be seen from Table \ref{MDparams}.   However, to what extent such change in the ground electronic state affects the disorder in the excitation energy is not clearly understood at this point.

As far as HB is concerned, $\alpha$-BChl of LH3 is expected to be equivalent to that of LH2.   However, there are some differences in their details. 
The HB of $\alpha$-BChl in LH3 is to Tyr41.   This conserves its relative intermolecular orientation close to that of the similarly hydrogen bound $\alpha$-BChl of LH2. However,  the torsional angle of the acetyl group of $\alpha$-BChl for LH3 is known to be different from that of LH2, as will be detailed below.

\subsection{Effect of acetyl group rotation}
\begin{figure}[htbp]
\begin{center}
(a)\makebox[3in]{ } \\
\includegraphics[width=3in]{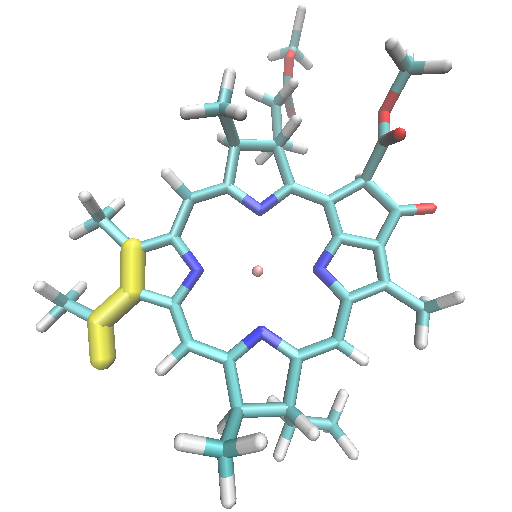}\\
(b)\makebox[3in]{ } \\
\includegraphics[width=\linewidth]{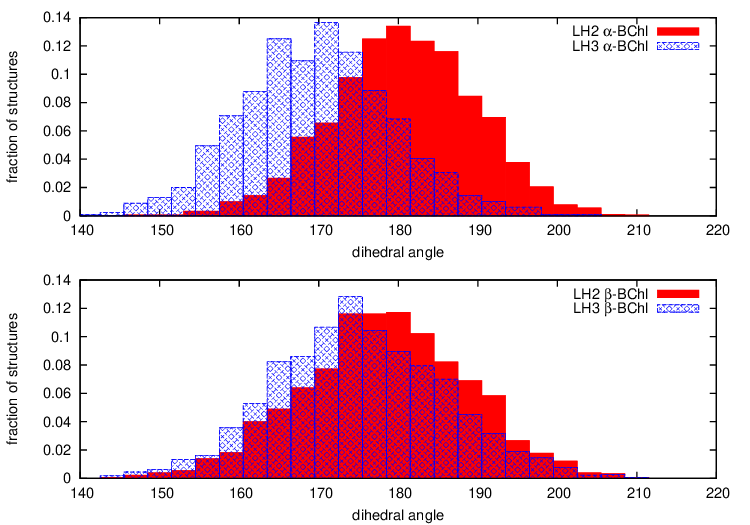}
\caption{(a) Structure of a BChl with phytyl chain remove.  The atoms ${\rm C2}$, ${\rm C3}$, ${\rm C3^1}$, and ${\rm O3^1}$ defining the dihedral (torsional) angle are highlighted as yellow. (b) Acetyl group dihedral angle distributions of BChls for LH2 (red) and LH3(blue).  The upper panel is for $\alpha$-BChls and the lower panel is for $\beta$-BChls.}
\label{dihedralDist}
\end{center}
\end{figure}

The acetyl group torsional angle is represented by the dihedral angle formed by atoms C2, C3, C3$^1$, and O3$^1$ (IUPAC nomenclature) of BChl (see Fig. \ref{dihedralDist} (a)).   The distribution of this angle sampled from the second phase MD simulations is shown in Fig. \ref{dihedralDist} (b) for the BChls from the B850 (LH2) and B820 (LH3) units.  The average values of these angles for $\beta$-BChls (including the error ranges estimated from the standard deviations) are respectively 179$^\circ$ $\pm$ 10.4$^\circ$ for LH2 and 176$^\circ$ $\pm$ 10.7$^\circ$ for LH3.  Thus, there is virtually no difference between average acetyl group torsional angles of $\beta$-BChls of LH2 and LH3, despite difference in their HB characteristics.  On the other hand, a noticeable disparity can be seen for $\alpha$-BChls.  For LH2, the torsional angle is 182$^\circ$ $\pm$ 8.9$^\circ$.  For LH3, it is  171$^\circ$ $\pm$ 9.4$^\circ$.  However, the resulting difference in the average angle of about 10$^\circ$ is still significantly less than what is suggested from the crystal structure data.\cite{mcluskey-biochemistry40}

We have investigated the dependence of the $Q_y$ excitation energies of $\alpha$-BChls  on the acetyl group rotation using the TD-DFT method.  For $\alpha$-BChls in the B850 unit, configurations with torsional angles near the mean values were used.  However, for $\alpha$-BChls in the B820 unit, we have used a structure with a torsional angle 156$^\circ$, which is different from the average value for LH2 by about ${\rm 26^o}$ and amounts to an extreme upper bound of the distribution observed from the MD simulation.   All the structures were further optimized  by the DFT method on the ground electronic state with four different functionals, M06-2X,\cite{zhao-tca120} LC $\omega$-PBE,\cite{vydrov-jcp125} CAM-B3LYP,\cite{yanai-cpl393} and $\omega$BD-97XD.\cite{chai-pccp10}  Not only the BChl, but also protein residue serving as the HB partner and the coordinating histidine group were included in this calculation. Table \ref{QMcalc} summarizes $Q_y$ excitation energies for $\alpha$-BChls calculated by the TD-DFT method.   All the calculations were performed by the Gaussian 09 software package,\cite{g09-d} employing the 6-31G(d,p) basis. 

As can be seen from Table \ref{QMcalc} and has been reported in our previous work,\cite{jang-jpcl6} the excitation energies of BChls differ significantly depending on the choice of the functional. On the other hand, the differences of the excitation energies of $\alpha$-BChls from LH2 and LH3 remain small and relatively insensitive to the functional.  Among those tested, the maximum difference was about ${\rm 166\ cm^{-1}}$ for LC $\omega$-PBE.  For all three others,  the differences are less than ${\rm 15 \ cm^{-1}}$.   These results suggest that the excitation energies of $\alpha$-BChls from LH2 and LH3 are not affected significantly by their differences in the torsional angle.  Note that these data are for rather extreme value of the torsional angle (more than twice the average value observed from MD simulations) for the latter.  For smaller torsional angles that are dominantly observed from the MD simulation, the differences in excitation energies are expected to be even smaller.  Thus, our calculation suggests that the rotation of the acetyl group is not likely to cause significant blue shift in the excitation energy of the $\alpha$-BChl of LH3.

Of course, there still remains possibility that the electronic correlation effect missing at the level of our TD-DFT calculation can be responsible for the shift.  However, we believe this is unlikely considering the consistency of the results across different functionals.  On the other hand, since it is known that TD-DFT is not capable of reproducing the wide range of variation in excitation energies of BChls due to subtle differences in interactions with overall protein environments,\cite{higashi-jpcb118} the possibility that other minor differences in the BChl-protein interactions can add up to significant blue shift remains plausible.  However, considering the non-specific nature of these interactions, they are expected to have similar influences on both the excitation energies of $\alpha$- and $\beta$-BChls.  Further discussion and test of of this possibility are provided in the next section.  

\begin{table}[htbp]
   \centering
   \caption{Excitation energies  (in eV) of $Q_y$ transitions for $\alpha$-BChls calculated by the TD-DFT method for structures optimized by the ground state DFT employing the same functional. For all the calculations, 6-31G(d,p) basis set was used.}
   \begin{tabular}{@{} c|ccc @{}} 
    \hline
    \hline   
     		& \makebox[1in]{$\alpha$-BChl in LH2}	& \makebox[1in]{$\alpha$-BChl in LH3} &(${\rm cm^{-1}}$)	\\
		\hline
M06-2X	& 14,532	& 14,545	&\\
LC $\omega$-PBE	& 11,635	& 11,469 &	\\
CAM-B3LYP	& 13,977	& 13,982&	\\
$\omega$ BD-97XD	& 13,521	& 13,507&	\\
      \hline
      \hline
   \end{tabular}
   \label{QMcalc}
\end{table}

\section{Discussion}
We have reported here comparative simulation and computational studies of both LH2 and LH3 complexes at room temperature and physiological condition, and provided a comprehensive set of data necessary for constructing  exciton-bath models for both complexes.  All the structural information obtained from our calculations/simulations can be compactly expressed by Eqs. (\ref{Rxyz}) and (\ref{muxyz}), along with parameters listed in Table \ref{geometry}.  All the energetic information in the single exciton space (without disorder) is represented by Eq. (\ref{eq:h_e0}) and Table \ref{MDparams}.  

The implications of our computational studies can be summarized  as follows:
\begin{itemize}
\item The $\alpha$-BChls of {\it in vivo} LH2 and LH3 complexes have difference in their torsional angles of acetyl groups by $\sim 10^o$, due to different local environments.  This difference observed from MD simulation data is significantly less than what was previously assumed based on the X-ray crystal structural data.  In addition, according to TD-DFT calculations, the observed change in torsional angle does not influence the $Q_y$ excitation energies of BChls significantly.     
\item The $\beta$-BChls observed in our MD simulation of  {\it in vivo} LH3 are missing HBs, which confirms what was expected from the X-ray crystal structure data.  According to our earlier computational study,\cite{jang-jpcl6} this suggests that the excitation energy of the $\beta$-BChl of LH3 has about ${\rm 500 \ cm^{-1}}$ higher value than that of LH2. 
\item The electronic couplings between BChls in LH3 are similar to those in LH2 except for two cases.  One is $J_{\alpha\beta}(1)$, the coupling between the nearest neighbor $\alpha$ and $\beta$-BChls of adjacent protomers, and the other is  $J_{\beta\gamma}(0)$, the coupling between $\beta$ and $\gamma$-BChls within the same protomer (see Table \ref{MDparams}).   However, although these changes are about or more than 50 \%, their effects on the spectroscopic features of the ensemble are expected to be rather minor.  This is because their absolute values are still significantly smaller than the typical values of energetic disorder of BChls generating the inhomogeneity of the ensemble lineshape. 

\item Our MD simulation results suggest that the spectral densities of the baths for LH2 and LH3 are virtually the same. 
\end{itemize}

Given that the features of LH3 as summarized above are the only distinctive ones,  one may expect that the lineshape of LH3 is similar to that of LH2 with single mutation by Fowler {\it et al.},\cite{fowler-nature355} where only one of the two HBs in each B850 protomer unit is broken.   However, the actual spectral position of the B820 band in LH3 is similar to that of LH2 with double mutations by Fowler {\it et al.}\cite{fowler-nature355}  

In order to confirm our assessment given above, we calculated ensemble absorption lineshapes employing the parameters determined from our simulation/computation and also from the fitting to experimental line shapes at room temperature.\cite{chmeliov-jpcb117}  A simple approximation that considers only the diagonal terms of the exciton-bath coupling in the basis of exciton Hamiltonian was used.  Although we have determined new spectral densities based on MD simulations here, they need further refinement such as including intra-molecular vibrational terms and further examination of convergence.  Thus, for the calculation of the line shape, we used the model spectral density developed before,\cite{jang-jpcb115} which is much more convenient to use.  In the definition of Eq. (\ref{eq:spd_sn}), this has the following form: 
\ben
{\mathcal J}_{s_n}(\omega)&=& \pi\hbar \eta_{s} \left (\gamma_1 \omega e^{-\omega/\omega_{c,1}}+\gamma_2\frac{\omega^2}{\omega_{c,2}} e^{-\omega/\omega_{c,2}}\right .\nonumber \\
&&\left .+\gamma_3\frac{\omega^3}{\omega_{c,3}^2}e^{-\omega/\omega_{c,3}}\right) \ ,
\een
where $\gamma_1=0.22$, $\gamma_2=0.78$, $\gamma_3=0.31$, $\hbar\omega_{c,1}=170\ {\rm cm^{-1}}$, $\hbar\omega_{c,2}=34\ {\rm cm^{-1}}$, and $\hbar\omega_{c,3}=69\ {\rm cm^{-1}}$.   The overall strength $\eta_s$ is on the order of unity, and remains as an adjustable parameter.   While this spectral density lacks high frequency vibrational modes, it is expected to give reasonable representation  near major excitonic peaks and thus serves our purpose. 

\begin{table}
\caption{Energetic parameters of theoretical models for LH2 and LH3.  $\epsilon_\alpha$, $\epsilon_\beta$, and $\epsilon_\gamma$ are average excitation energies of $\alpha$, $\beta$, and $\gamma$ BChls.  $\sigma_\alpha$, $\sigma_\beta$, $\sigma_\gamma$, and $\sigma_g$ are standard deviations of the Gaussian disorder in the excitation energies of $\alpha$, $\beta$, and $\gamma$ BChls, and that in the ground state energy. }
\begin{tabular}{c|cccccccc}
\hline
\hline
&\makebox[.35in]{$\epsilon_\alpha$}&\makebox[.35in]{$\epsilon_\beta$}&\makebox[.35in]{$\epsilon_\gamma$}&\makebox[.25in]{$\sigma_\alpha$}&\makebox[.25in]{$\sigma_\beta$}&\makebox[.25in]{$\sigma_\gamma$}&\makebox[.2in]{$\sigma_g$}&(${\rm cm^{-1}}$)\\
\hline
Model A&12160&12160&12520&270&270&65&70&\\
Model B&12160&12660&12520&270&270&65&70&\\
Model C&12480&12980&12580&260&240&65&70&\\
\hline
\end{tabular}
\label{table:model_energies}
\end{table}
\ \vspace{.5in}\\

\begin{figure}
\ \vspace{.5in}\\
\includegraphics[width=0.7\linewidth]{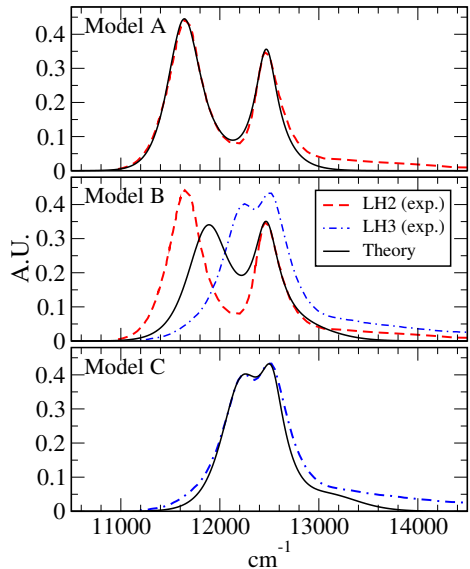}\hspace{.5in}\\
\caption{Theoretical  ensemble lineshapes for models  A, B, and C, each averaged over $10,000$ realizations of disorder, compared with experimental ensemble lineshapes\cite{chmeliov-jpcb117} for LH2 and LH3 complexes.}
\label{fig:lineshapes}
\end{figure}

After testing various values of as yet undetermined parameters, we have identified three representative models.  Model A corresponds to the parameters of LH2 in Tables \ref{geometry} and \ref{MDparams}, the data shown in the first row of Table \ref{table:model_energies}, $\eta_\alpha=\eta_\beta=1$, and $\eta_\gamma=1.1$.      The top panel of Fig. \ref{fig:lineshapes} compares the theoretical ensemble line shape of the model A averaged over $10,000$ realizations of disorder with an experimental line shape\cite{chmeliov-jpcb117} for LH2. There is some discrepancy in the blue side, which seems to be in part due to the lack of high frequency intramolecular vibrational modes and relative inaccuracy of the model spectral density for the B800 unit, as can be seen from Fig. \ref{specden}.  Otherwise, the agreement between theoretical and experimental lineshapes is satisfactory, which suggests that Model A is a reasonable representation of LH2. 

Model B correspond to the parameters of LH3 in Tables \ref{geometry} and \ref{MDparams}, the data shown in the second row of Table \ref{table:model_energies}, $\eta_\alpha=\eta_\beta=1$, and $\eta_\gamma=1.1$.  The values of parameters in Table \ref{table:model_energies} are the same as those in Model A except for the average excitation energy of $\beta$-BChl.   Thus, this model B represents the case where the excitation energies of $\beta$-BChls are shifted to the blue by $500\ {\rm cm^{-1}}$, while the excitation energies of $\alpha$-BChls remain the same.  The middle panel of Fig. \ref{fig:lineshapes} shows calculated ensemble line shape averaged over $10,000$ realizations of disorder, compared to experimental line shapes\cite{chmeliov-jpcb117} of both LH2 and LH3 complexes.   The clear difference between the theoretical line shape and the experimental line shape for LH3 points to the need for identifying an additional source of the blue shift.

A recent work by Chmeliov {\it et al.}\cite{chmeliov-jpcb117} suggests that detailed account of all BChl-protein interactions, although at molecular mechanics level, may be sufficient to account for the discrepancy.  This is interesting because it suggests that a collection of non-specific interactions with protein environments can result in significant change in the excitation energies of BChls.  Based on this observation and after a series of test calculations, we identified the Model C, which corresponds to the parameters of LH3 in Tables \ref{geometry} and \ref{MDparams}, the data in the third row of Table \ref{table:model_energies}, $\eta_\alpha=\eta_\beta=1$, and $\eta_\gamma=1.2$.   In this model, in addition to the blue shift in the excitation energy of $\beta$-BChls by $500\ {\rm cm^{-1}}$ compared to that of $\alpha$-BChl, there is also common blue shift of the excitation energies of both $\alpha$ and $\beta$ BChls by $320\ {\rm cm^{-1}}$, which represents net contribution of nonspecific interactions with all protein residues.  We have also made slight adjustment of other parameters as can be seen from Table \ref{table:model_energies}.  The bottom panel of Fig. \ref{fig:lineshapes} compares the theoretical line shape of this model, also averaged over $10,000$ realizations of disorder, with the experimental one for LH3.  Although there are some discrepancies in both the red and blue sides, this model reproduces the major feature of the experimental lineshape of LH3 fairly well, which suggests that the parameters of Model C serve as reasonable representation of LH3.  Further improvement of this may be possible by using  more accurate line shape theory and spectral densities.

\section{Conclusion} 

The exciton-bath Hamiltonian for both LH2 and LH3 complexes and the values of parameters determined here, which are summarized in Tables \ref{geometry} and \ref{MDparams}, can be used for simulating large scale exciton dynamics in the photosynthetic unit of purple bacteria.   We also made careful examination of molecular level characteristics that  differentiate the two complexes, and their relationship to their spectral features.  From the results of direct MD simulations,  we confirmed that only $\beta$-BChls of LH3 lack stable HB.  On the other hand, $\alpha$-BChls of LH3 maintain stable HB although their acetyl group torsional angles are more distorted from being planar than those for LH2.  

 According to TD-DFT calculations, we estimate that the lack of HB causes blue shift of the excitation energy by about $500\ {\rm cm^{-1}}$ but that the distortional of torsional angle of $\alpha$-BChl as observed does not cause significant change in the excitation energy.   These two characteristics alone are insufficient to explain the shift of the LH3 complex. However, when we introduced additional blue shift of both $\alpha$- and $\beta$-BChls by ${\rm 320\ cm^{-1}}$, we find that the resulting theoretical line shape agrees reasonably well experimental ones.   The source of this common blue shift is not clear at this point, but may be due to net contribution of slightly different protein environments around BChls in the LH3 complex.   Further investigation of this possibility requires larger scale {\it ab initio} calculations including BChl-protein interactions explicitly.

\acknowledgements
This work was mainly supported by the Office of Basic Energy Sciences, Department of Energy (DE-SC0001393), and in part by the National Science Foundation  (CHE-1362926) for the development of computational methods involved.  SJ also acknowledge the support of the Camille Dreyfus Teacher Scholar Award during the early stage of this work.   Authors acknowledge computational support by the CUNY High Performance Computing Center, Queens College Center for Computational Infrastructure for the Sciences, and the Center for Functional Nanomaterials at Brookhaven National Laboratory.

\providecommand{\latin}[1]{#1}
\providecommand*\mcitethebibliography{\thebibliography}
\csname @ifundefined\endcsname{endmcitethebibliography}
  {\let\endmcitethebibliography\endthebibliography}{}

\end{document}